\documentclass[aps,prb]{revtex4}
\usepackage{amsmath,amssymb}
\usepackage{graphicx}
\usepackage{dcolumn}
\usepackage{bm}
\usepackage{color}
\usepackage{relsize}
\newcommand{\mbb}{\mathbb}
\newcommand{\mc}{\mathcal}

\newcommand{\tet}{\texttt}
\newcommand{\pr}{\partial}

\begin{document}

\title{Tunneling conductivity fast modulated by optically-dressed electrons \\ in graphene and a dice lattice}


\author{Andrii Iurov$^{1}$\footnote{E-mail contact: aiurov@mec.cuny.edu, theorist.physics@gmail.com},  
Liubov Zhemchuzhna$^{1,2}$,
Godfrey Gumbs$^{2,3}$, 
Danhong Huang$^4$,
and 
Paula Fekete$^5$
}

\affiliation{
$^{1}$Department of Physics and Computer Science, Medgar Evers College of City University of New York, Brooklyn, NY 11225, USA\\ 
$^{2}$Department of Physics and Astronomy, Hunter College of the City University of New York, 695 Park Avenue, New York, New York 10065, USA\\ 
$^{3}$Donostia International Physics Center (DIPC), P de Manuel Lardizabal, 4, 20018 San Sebastian, Basque Country, Spain\\ 
$^{4}$US Air Force Research Laboratory, Space Vehicles Directorate, Kirtland Air Force Base, New Mexico 87117, USA\\ 
$^{5}$US Military Academy at West Point, 606 Thayer Road, West Point, New York 10996, USA
}

\date{\today}

\begin{abstract}
Based on the transmission coefficient of tunneling electrons, we have presented tunneling current and conductivity across a square-potential barrier for both graphene and $\alpha$-$\mc{T}_3$ lattices under a linearly-polarized off-resonant dressing field. The presence of such a dressing field introduces an anisotropy factor in the energy dispersion of tunneling electrons so that the cross section of a Dirac-cone appears as elliptical.  Consequently, the field-polarization controlled major axis of the ellipse will be misaligned with the normal direction of a barrier layer in the tunneling system, which exhibits an asymmetric Klein-paradox for an  off-normal-direction tunneling. The resulting tunneling current in this system is calculated by using a transmission coefficient and a longitudinal group velocity (different from a longitudinal momentum) of electrons.  By presenting numerically calculated tunneling conductivity modified by a laser dressing field, we demonstrate a significant enhancement of electrical conductivity by external laser-field intensity, which is expected to be crucial in application of ultrafast optical modulation of opto-electronic devices for photo-detection and fiber-optic communication. 
\end{abstract}

\maketitle

\section{Introduction} 
\label{sec1}

The  $\alpha-\mc{T}_3$  model represents an innovative and unusual type of Dirac materials in which their atomic structure leads to a pseudospin-1 Dirac-Weyl model Hamiltonian for electrons and an additional energy band which is completely flat and symmetric with respect to the valence and conduction bands corresponding to the upper and lower parts of the Dirac cone in graphene.\,\cite{leykam2018artificial,li2018realization,cunha2021band,wang2011nearly,dora2011lattice} This type of energy bandstructure results in a variety of truly unique and previously unknown electronic,\,\cite{illes2017klein}  topological,\,\cite{dey2019floquet}  collective,\,\cite{malcolm2016frequency,iurov2020many}  optical \,\cite{illes2015hall,bryenton2018optical} and magnetic properties \,\cite{illes2016magnetic,piechon2015tunable,raoux2014dia,malcolm2014magneto} of $\alpha-\mc{T}_3$  model.

\medskip
\par
The atomic composition of an $\alpha-\mc{T}_3$ lattice consists of a regular graphene-like honeycomb lattice with additional fermionic atoms at the center of each hexagon, which are referred to as hub atoms in contrast to the rim atoms located at each of six hexagonal vertices. Here, the ratio between the hub-rim and rim-rim electron hopping coefficients defines a parameter $\alpha$ with its limiting values $\alpha=0$ for graphene and $\alpha=1$ for a dice lattice\,\cite{illes2017properties, moller2012correlated}.   
One of the first realizations of a crystal lattice with a non-dispersive flat energy band in derivations of the bandstructure dates back to nearly a decade ago\,\cite{shen2010single, bercioux2011topology}. At the present time, there has already been copious and convincing evidence for experimental confirmation and successful-fabrication efforts in producing different types of $\alpha-\mc{T}_3$ materials, in which the most well-known cases include a three-layer structure of SrTiO$_3$/SrIrO$_3$/SrTiO$_3$ lattices with a cubic crystal symmetry,\,\cite{wang2011nearly} Lieb and Kagome lattices with additional atoms at the edges\,\cite{slot2017experimental, guo2009topological, xue2019acoustic, yin2019negative,lin2018flatbands,kang2020topological}, and a number of other planar  materials\,\cite{betancur2017perfect,milicevic2019type,leykam2018perspective} as well.

\medskip
\par
On the other hand, Floquet engineering \,\cite{holthaus2015floquet, oka2019floquet, goldman2014periodically, bukov2015universal} appears as a crucial tool for the modification of principal electronic properties of a planar structure\,\cite{usaj2014irradiated, perez2015hierarchy} or the surface states in a three-dimensional material \,\cite{kyriienko2019optically,calvo2015floquet} by employing an off-resonance dressing field with its frequency substantially higher than the characteristic energies of a system, such as the Fermi energy. Encouraged by rapid technological advancements in intense lasers and microwave field sources, lots of new field-induced phenomena in recently discovered low-dimensional structures, especially Dirac materials, have been demonstrated both theoretically and experimentally.\,\cite{cheng2019observation,tokman2019second}
Physically, a strong off-resonance field applied to a two-dimensional material enables the appearance of so-called dressed states, i.e., a quantum-mechanically composite involving both irradiated electrons and the applied field. Such a quantum composite acquires tunable electronic properties controlled by the strength of an external laser field.\,\cite{kyriienko2017floquet} 
Moreover, the laser-dressing field can also induce new or significantly modify existing fundamental quantum effects in the target  system.\,\cite{iorsh2017optically,iorsh2021optically,morina2015transport,kibis2020light}  

\medskip
\par
The type as well as major dynamical properties of these dressed states are mainly manipulated by the polarization of external dressing field, e.g., circular-polarized light is known for opening or varying an existing energy gap,\,\cite{iurov2017exchange,kibis2010metal,iurov2020quantum,kibis2017all} while linearly-polarized field can induce an anisotropy\,\cite{iurov2017exploring} in optical response of irradiated materials. The theoretical study of laser-induced anisotropy in opto-electronic properties of graphene and a dice lattice is the focus of current research. Specifically, we will concentrate on the tunneling-transport properties of a system with light-induced tunable anisotropy of its electronic states.\,\cite{ourpeculiar,kristinsson2016control}

\medskip
\par
The electronic property of highly anisotropic Dirac fermions was first predicted theoretically by using the first-principles computations
and then confirmed experimentally\,\cite{feng2018discovery} based on angle-resolved photoemission spectroscopy of epitaxial graphene modulated by an island superlattice,\,\cite{rusponi2010highly} a 
Bi-square net of SrMnBi$_2$,\,\cite{park2011anisotropic} an organic conductor,\,\cite{hirata2016observation} and a B$_2$S honeycomb monolayer\,\cite{zhao2018new}. Such kind of materials has revealed  unique features of an exceptional Dirac-material family as long as engineering a local directional asymmetry\,\cite{rusponi2010highly} can be introduced into the system.
Since these materials with anisotropic dispersions are available and implemented within an electronic device, most of their electronic and collective properties have been 
addressed thoroughly, including plasmons \,\cite{hayn2021plasmons,politano20183d} whose frequency depends on the direction of wave vector, as well as angle-dependent screening and transport.\,\cite{trescher2015quantum,zhang2019borophosphene} Meanwhile, specific attention has also been put onto materials with a tilted Dirac cone and having either a zero or finite bandgap, which exhibit a unique out-of-plane anisotropy\,\cite{sari2014magnetoplasmons,tan2021anisotropic,trescher2015quantum,lu2016tilted,jalali2018tilt} and even include recently fabricated 8-Pmmn borophene 
as well as single-element materials presenting two ionic sublattices.\,\cite{lopez2016electronic,sadhukhan2017anisotropic,zhou2020anomalous,wang2019band,ibarra2019dynamical,champo2019metal}

\medskip
\par
Historically, Klein paradox, a complete and unimpeded tunneling of incoming charged carriers through a square-potential barrier with arbitrary height and width, has become one subject unique to all gapless Dirac cone materials.\,\cite{katsnelson2006chiral} This phenomena is attributed to the relativistic type of graphene Hamiltonian and the existence of chirality in its electronic wave functions. 
Interestingly, Klein tunneling was demonstrated extensively in graphene, a dice lattice and all types of interpolating $\alpha-\mc{T}_3$ even under a tilted potential barrier.\,\cite{illes2017klein,urban2011barrier,ye2020quantum,betancur2017super,mandhour2020klein} Furthermore, Klein paradox is found persisting for an anisotropic Dirac cone. However, 
in this case, it occurs only at a finite angle of incidence and is referred to as an asymmetric Klein tunneling.\,\cite{sonin2009effect,weekes2021generalized}
Technically, however, various types of potential barriers, e.g., inhomogeneous and non-uniform spatial profiles of a potential and junctions, could be easily realized in graphene or a nanoscale-width nanoribbon by introducing a spatially-distributed gate voltage.\,\cite{sengupta2008tuning,cho2007gate, moon2013lateral} 
As a part of graphene-based optoelectronic device,\,\cite{anwar2020interplay,britnell2013resonant,AAnwar2021}
studying electron conductance through these different barrier arrangements, as well as revealing involved physics mechanism for ballistic transport,\,\cite{masir2009tunneling,iurov2013photon,bhattacharjee2006tunneling,stauber2008conductivity} are paramount for quantitatively predicting the current level and characterizing  
the performance of an electric switch.

\medskip
\par
The remaining part of this paper is organized as follows. In Sec.~\ref{sec2}, we review some of important properties of optically modulated Dirac electrons with tunable-elliptical energy dispersions of electrons irradiated by linearly-polarized light, including the calculation of transmission of electrons and demonstrating the so-called asymmetric Klein tunneling associated with a non-head-on electron incidence. Based on computed transmission results, we arrive at an expression for the tunneling conductivity of electrons over a square  potential barrier in Sec.~\ref{sec3} and provide a detailed investigation on how this conductivity relies on the crucial material and external irradiation parameters, e.g., the irradiation-induced anisotropy factor $a(\lambda_0)$ in the energy dispersion of electrons and the misalignment angle between the direction of light polarization and the normal direction of a barrier layer. These new features are not known from previously considered cases for isotropic particles. The final conclusions are drawn in Sec.~\ref{sec4}.

\section{Asymmetric Klein tunneling resulting  from elliptical Dirac dispersions}
\label{sec2}

The energy dispersion of electrons in a crystal usually appears spherical around a highly-symmetric valley such as the $\Gamma$ point, within the first Brillouin zone. However, in the presence of strong laser-electron interaction, this energy dispersion is modified and becomes anisotropic, e.g., an elliptical one with a long-axis parallel to the polarization direction of an incident laser field. If the long-axis of the elliptical energy dispersion appears misaligned with the normal direction of the square-barrier layer in a tunneling structure, we expect to find that the tunneling conductivity
will vary with this misalignment angle. Therefore, the first step for computing tunneling and conductivity in our system is to derive  anisotropic dressed states of electrons under a  linearly-polarized dressing field and to full understand how this anisotropy depends on the intensity of an applied optical field in different Dirac materials.  Even though we only consider in this paper graphene ($\alpha\to 0$) and a dice lattice ($\alpha\to 1$) as two extreme limiting cases for the $\alpha-\mc{T}_3$ model, it would be useful to present a comprehensive description for the induced anisotropy by a linearly-polarized dressing field in all $\alpha-\mc{T}_3$ materials.  

\medskip
\par
For a two-dimensional (2D) $\alpha-\mc{T}_3$ material, the low-energy Hamiltonian takes the form

\begin{equation}
\label{H0}
\hat{\mc{H}}_{\alpha} (\mbox{\boldmath$k$} \, \vert \, \tau, \phi) = \hbar v_F \,
\left[
\begin{array}{ccc}
0 & k^\tau_- \cos \phi &  0 \\
0 & 0 & k^\tau_-\sin \phi \\
0 & 0 & 0
\end{array}
\right]+ H.c.\ ,  
\end{equation}
where the geometry phase $\phi=\tan^{-1}\alpha$ is directly related  to the relative hopping parameter $\alpha$, and $k^\tau_\pm=k^\tau_x\pm ik^\tau_y$ are obtained from components of the electron wave vector $\mbox{\boldmath$k$}=\{k_x,k_y\}$.
Once an optical dressing field {\it with a linear polarization} is applied, the components of $\mbox{\boldmath$k$}$ introduced in the Hamiltonian  
in Eq.\,\eqref{H0} are modified according to a canonical substitution, i.e. $k_{i} \to  k_{i} - e A_i^{(L)}(t)$, where the transient  
vector potential $\mbox{\boldmath$A$}^{(L)}(t)$ of this linearly-polarized light is given by  

\begin{equation}
\label{linA}
\mbox{\boldmath$A$}^{(L)}(t) = 
\left[  \begin{array}{c}
          A^{(L)}_x (t) \\
          A^{(L)}_y (t)
        \end{array}
\right] = \frac{E_0}{\omega} \left[
\begin{array}{c}
\cos\beta\\
\sin\beta
\end{array}
\right] \, \cos (\omega t ) \ ,
\end{equation}
with  $E_0$ representing the electric field amplitude, $\omega$ is the frequency of light, and $\beta$ stands for the polarization angle of light made with the $x$ axis. Moreover, in the presence  of both the dressing field and an electrostatic finite-width square barrier $V(x)=V_0\,\Theta(x)\,\Theta(W_B-x)$, the non-interacting Hamiltonian in Eq.\,\eqref{H0}   becomes

\begin{equation}
\label{HamG}
\hat{\mbb{H}}_0^\tau (\phi \, \vert \, x, y) = - v_F \, \hat{\mbox{\boldmath$S$}}(\phi) \cdot \left[ i \hbar \mbox{\boldmath$\nabla$}_\tau + e\mbox{\boldmath$A$}^{(L)}(t) \right] + V(x) \ ,
\end{equation}
where $\Theta(x)$ is the Heaviside step function, $V_0$ and $W_B$ are barrier height and thickness respectively, $\mbox{\boldmath$\nabla$}(\tau)\equiv \{ \tau \pr/\pr x,\,\pr/\pr y \}$, and the two $\phi$-dependent matrices $\hat{\mbox{\boldmath$S$}}(\phi)= \{\hat{S}_x(\phi),\hat{S}_y(\phi)\}$ are defined as 

\begin{equation}
\label{Sxp}
\hat{S}_x(\phi) = \left[
\begin{array}{ccc}
 0 & \cos \phi & 0 \\
 \cos \phi & 0 & \sin \phi \\
 0 & \sin \phi & 0
\end{array}
\right] \ ,
\end{equation}

\begin{equation}
\label{Syp}
\hat{S}_y(\phi) = i \,\left[
\begin{array}{ccc}
 0 & -\cos \phi & 0 \\
 \cos \phi & 0 & -\sin \phi \\
 0 & \sin \phi & 0
\end{array}
\right] \ .
\end{equation} 
Here, even if the potential $V(x)$ only remains piecewise-constant, the translational symmetry of the system is still maintained and, therefore, we can simplify our eigenvalue equation 
using $\{\pr/ \pr x, \pr/ \pr y \} \rightarrow i \, \{k_x, k_y\}$ and assuming the wave function in the form of $\Psi(x,y) \backsim \tet{e}^{i k_x \, x} \tet{e}^{i k_y \, y}$.   Consequently, the energy dispersions obtained from the Hamiltonian in Eq.\,\eqref{HamG} for the dressed states keep the flat band 
$\mc{E}_\alpha^{\gamma = 0} (\lambda_0, \mbox{\boldmath$k$}) = 0$ under the dressing field, while two anisotropic Dirac-cone dispersions are given by  

\begin{equation}
\label{linD}
\mc{E}_\alpha^{\,\gamma = \pm 1} (\lambda_0, \mbox{\boldmath$k$}) =
 \pm \hbar v_F \sqrt{k_x^2+[a_\alpha(\lambda_0)\,k_y]^2} \equiv
 \pm \hbar v_F k \,  f_\alpha(\theta_{\bf k},\lambda_0)\ , 
\end{equation}
which appear as an ellipse in the $(x,y)$-plane. Here, the angular factor in Eq.\,\eqref{linD} is calculated as

\begin{equation}
\label{TA}
f_\alpha(\theta_{\bf k},\lambda_0)  = 
 \cos^2 \theta_{\bf k} + \left[
 J^2_0 (2 \lambda_0) \cos^2(2 \phi) + 
 J^2_0 (\lambda_0) \sin^2(2 \phi)
 \right] \,\sin^2 \theta_{\bf k}\ ,
\end{equation} 
where $J_0 (x)$ stands for the zeroth-order Bessel function of the first kind. In addition, we know that, for fixed $\theta_{\bf k}$, the anisotropic factor $f(\theta_{\bf k},\lambda_0)$ in Eq.\,\eqref{TA} depends on $\alpha$ or phase $\phi$ and 
it reaches a maximum for $\phi=0$ (graphene) but a minimum for $\phi= \pi/4$ (dice lattice). In particular, the anisotropy employed in Eq.\,\eqref{linD} takes the form 
 
\begin{equation}
\label{aphi}
a_\alpha(\lambda_0) = 1 - \frac{\lambda_0^2}{8} \left[ 5 + 3 \cos (4 \phi) \,\right]\ ,
\end{equation}
which is always less than one (isotropy with $f_\alpha(\theta_{\bf k},\lambda_0)\equiv 1$) if the irradiation intensity $\lambda_0\propto E_0^2\neq 0$ and reveals how the strength of a dressing field modifies the interaction property    
of $\alpha-\mc{T}_3$ materials. For example, we find from Eq.\,\eqref{aphi} a relatively larger anisotropy with $a_0(\lambda_0) = 1 - \lambda_0^2$ for graphene but a smaller anisotropy with $a_1(\lambda_0) = 1 - \lambda_0^2/4$ for dice lattice. 
\medskip

\begin{figure} 
\centering
\includegraphics[width=0.8\textwidth]{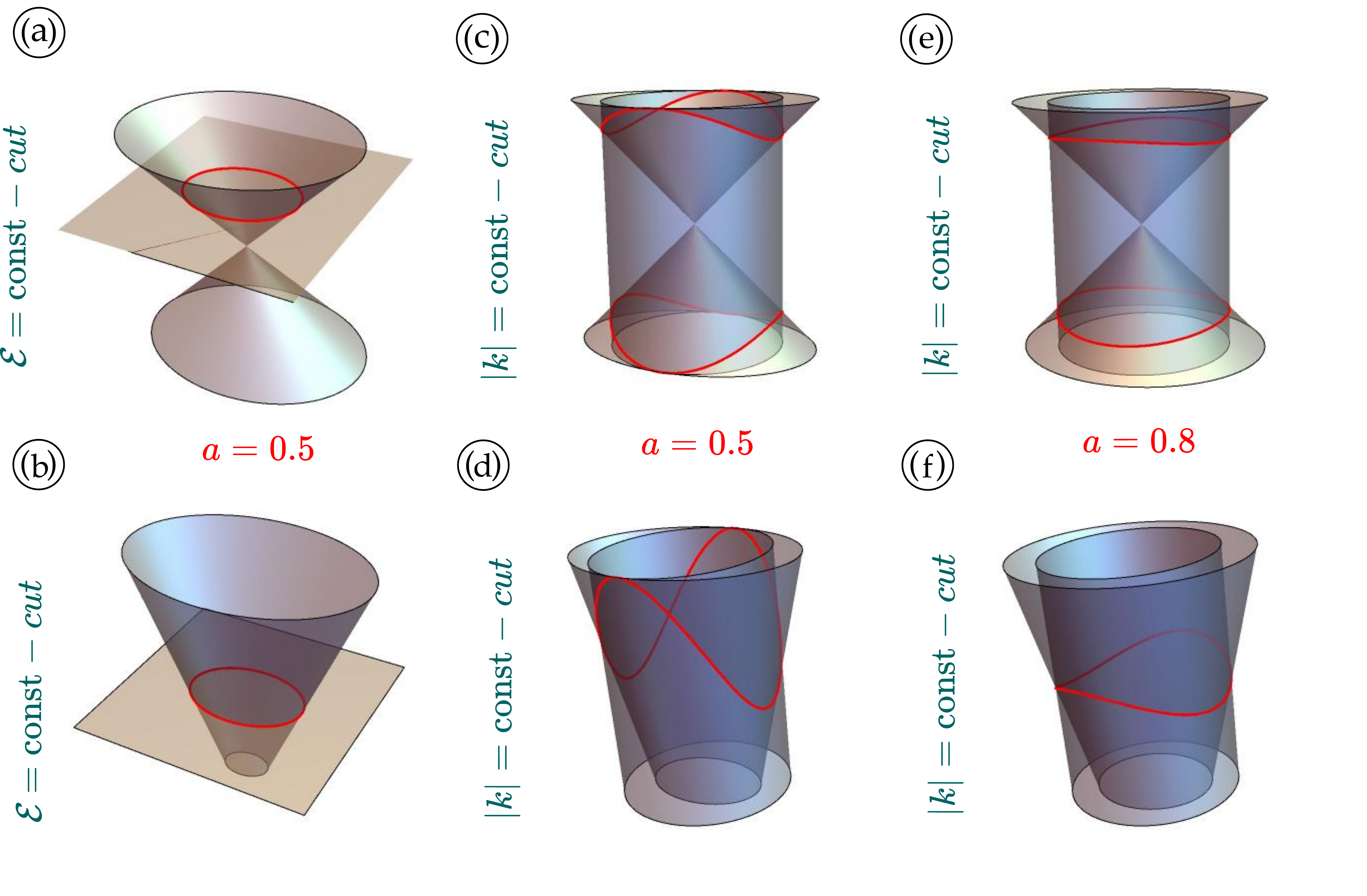}
\caption{(Color online) A horizontal (constant energy $\mc{E}_0$) cut, as well as a vertical (constant wave-vector magnitude $\vert\mbox{\boldmath$k$}\vert$) cut, 
of anisotropic Dirac-cone dispersions for both electron and hole states ($\gamma = \pm 1$), which are represented by intersection curves of 
the dispersion cones with either a horizontal plane or an upright cylinder. Panels $(a)$-$(d)$ describe the situations with $a_\lambda=0.5$ for 
a larger anisotropy, while plots $(e)$ and $(f)$ correspond to $a_\lambda=0.8$ for a smaller anisotropy. Here, the lower panels $(b)$, $(d)$, $(f)$ are 
zoom-in views for the energy range $0.2 < \mc{E}_0/E^*_F < 1.0$ with Fermi energy $E^*_F$ for a host material.}
\label{FIG:1}
\end{figure}

Figure\,\ref{FIG:1}\,\footnote{A brief and informative description and a recipe for finding an intersection curve for two given surfaces using Wolfram Mathematica similar to what was used here could be found in https://community.wolfram.com/groups/-/m/t/177994.} displays both horizontal and vertical cuts of an anisotropic Dirac-cone in different energy scales.  
As shown in Fig.\,\ref{FIG:1}, the constant-energy or horizontal cut of the anisotropic Dirac-cone presented in Eq.\,\eqref{linD} turns into an ellipse with a shortened $k_y$ semi-axis. 
This ellipse is inscribed by a $\vert\mbox{\boldmath$k$}\vert$-circle which includes all $\mbox{\boldmath$k$}_\lambda$ wave vector satisfying Eq.\,\eqref{linD} and  
$\vert\mbox{\boldmath$k$}_\lambda\vert\leq\vert\mbox{\boldmath$k$}\vert$. The constant $\vert\mbox{\boldmath$k$}\vert$-cut by an upright cylinder, on the other hand, appears as a much more sophisticated three-dimensional curve  
which acquires its maximum values on the $k_y$ axis as $\theta_{\bf k} = \pi/2$. 
\medskip

\begin{figure} 
\centering
\includegraphics[width=0.75\textwidth]{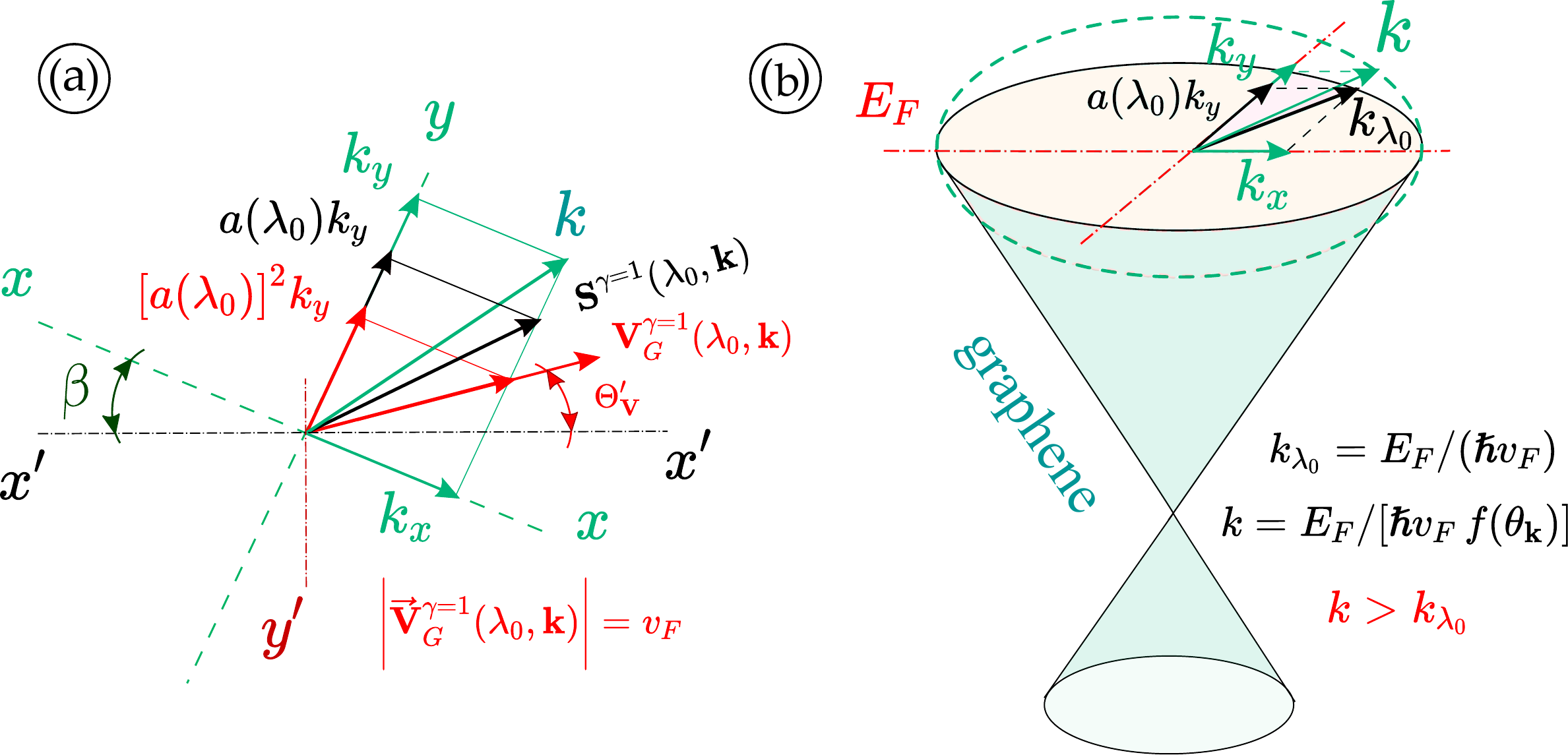}
\caption{(Color online) A schematic graph for determining the momentum of transmitted $\mbox{\boldmath$k$}^{(2)}$ and 
reflected $\mbox{\boldmath$k$}^{(r)}$ electrons, as well as the group-velocity vector $\mbox{\boldmath$V$}_{G}^{\gamma=1} (\lambda_0, \mbox{\boldmath$k$})$
in the barrier region characterized by a barrier height $V_0$. These quantities could be computed by the known energy difference 
$\mc{E}_0 - V_0$ in the barrier region and the fact that the transverse electron momentum $k'_y$ remains conserved in this system.} 
\label{FIG:2}
\end{figure}

Corresponding to energy dispersion in Eq.\,\eqref{linD}, the wave functions for the electron and hole states (i.e., conduction and valence bands with $\gamma=\pm 1$) in a dice lattice ($\alpha=1$)
are given as 

\begin{equation}
\label{wfdice1F}
\Psi_1^{\gamma = \pm 1} (\tau, \lambda_0, \mbox{\boldmath$k$}) = \frac{1}{4} \, 
\left[
\begin{array}{c}
\tet{e}^{- i \Theta^{(1)}_{\bf S} (\tau, {\bf k} \, \vert \, \lambda_0)} \\
\sqrt{2}  \, \gamma  \\
\tet{e}^{+ i \Theta^{(1)}_{\bf S} (\tau, {\bf k} \, \vert \, \lambda_0)}
\end{array}
\right] \ ,
\end{equation}  
where the phase factor, as defined in Fig.\,\ref{FIG:2}, is calculated from 

\begin{equation}
\label{ts0}
\Theta^{(1)}_{\bf S} (\tau, \mbox{\boldmath$k$} \, \vert \, \lambda_0) = \tan^{-1} \left[  
\left(\frac{\tau k_y}{k_x}\right) a_1(\lambda_0) \,
\right] =  \tan^{-1} \left[\tau a_1(\lambda_0)\tan \theta_{\bf k}\right]\ . 
\end{equation} 
Moreover, the wave function attributed to the flat band ($\gamma=0$) is 
 
\begin{equation}
\label{wfdice0}
\Psi_1^{\gamma=0} (\tau, \lambda_0, \mbox{\boldmath$k$}) = 
 \left[
\begin{array}{c}
\tet{e}^{- i \Theta^{(1)}_{\bf S}(\tau, {\bf k} \, \vert \, \lambda_0)} \\
0 \\
- \tet{e}^{+ i \Theta^{(1)}_{\bf S} (\tau, {\bf k} \, \vert \, \lambda_0)}
\end{array}
\right]
=
\frac{1}{k_\lambda} \, 
\left[
\begin{array}{c}
k_x - i \tau a_1(\lambda_0) k_y \\
0 \\
-k_x - i \tau a_1(\lambda_0) k_y   
\end{array}
\right] \ ,
\end{equation}
where $k_\lambda=\sqrt{k_x^2+a_1^2(\lambda_0)k_y^2}$ represents a measure for the kinetic energy of incident particles.
\medskip

\begin{figure} 
\centering
\includegraphics[width=0.75\textwidth]{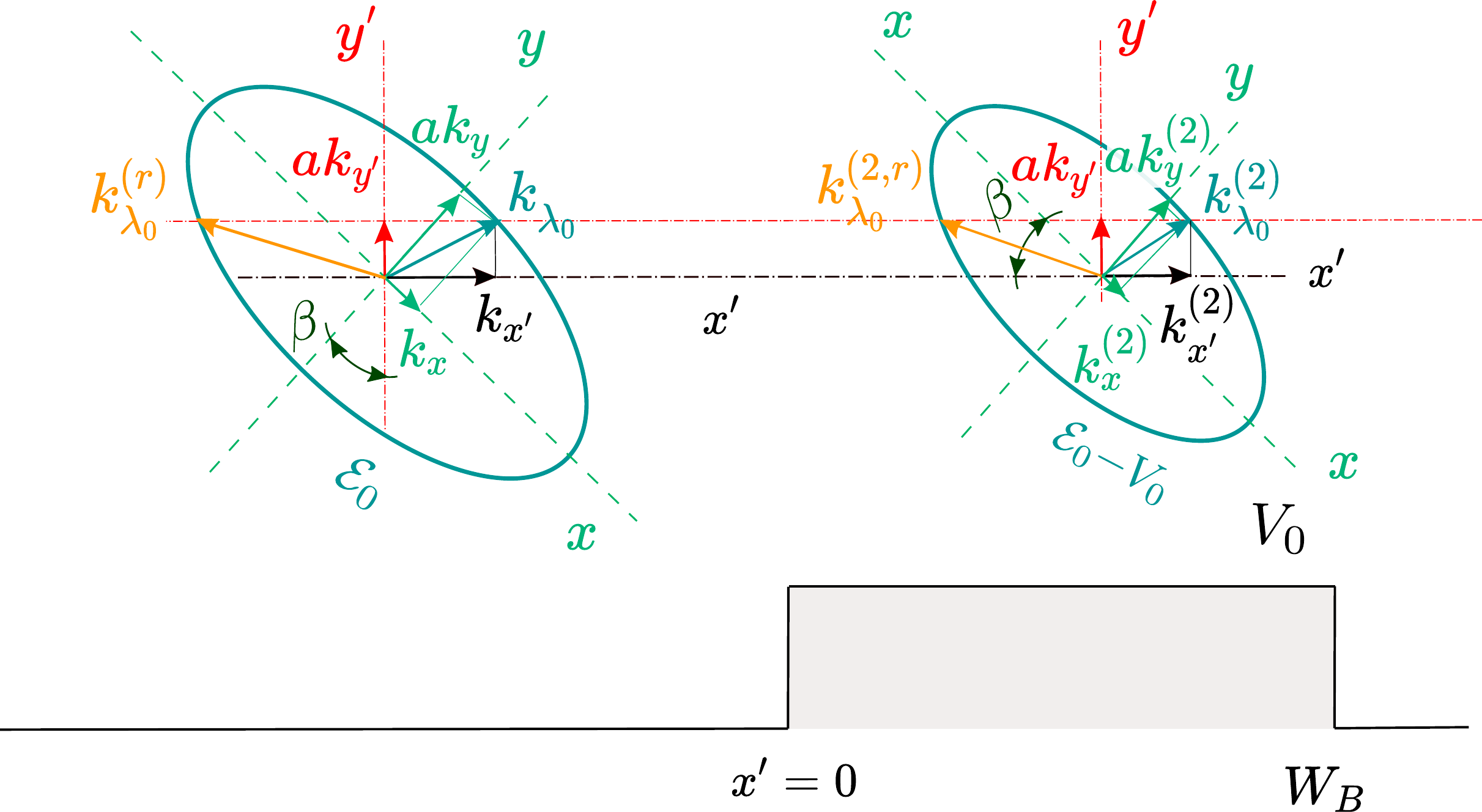}
\caption{(Color online) $({\rm Left})$ The relationships between the electron wave vector $\mbox{\boldmath$k$}$, spinor 
$\mbox{\boldmath$S$}^{\gamma=1} (\lambda_0, \mbox{\boldmath$k$})$ and group-velocity vector $\mbox{\boldmath$V$}_{G}^{\gamma=1} (\lambda_0, \mbox{\boldmath$k$})$, i.e., 
a visualization of Eq.\,\eqref{SVa}, outside the barrier layer. $({\rm Right})$ Determining the components of wave vector $\mbox{\boldmath$k$}$ within the barrier layer, corresponding to a given energy 
${\cal E}_0=E_F^*$ required for calculating tunneling conductivity in Eq.\,\eqref{sigma0}.}
\label{FIG:3}
\end{figure}

We further see from Fig.\,\ref{FIG:3} that the wave vector $\mbox{\boldmath$k$}$, the spinor vector $\mbox{\boldmath$S$}^{\gamma}(\lambda_0, \mbox{\boldmath$k$})$ and 
the group-velocity vector $\mbox{\boldmath$V$}_G^{\gamma}(\lambda_0, \mbox{\boldmath$k$})$ are different and not aligned to each other in the existence of a finite
anisotropy $a_1(\lambda_0) \neq 1$. Explicitly, $\mbox{\boldmath$S$}^{\gamma}(\lambda_0, \mbox{\boldmath$k$})$ and $\mbox{\boldmath$V$}_G^{\gamma}(\lambda_0, \mbox{\boldmath$k$})$ are defined as   

\begin{eqnarray}
\label{SV}
&& \mbox{\boldmath$S$}^{\gamma}(\lambda_0, \mbox{\boldmath$k$}) = \frac{\gamma}{\sqrt{k_x^2 + [a_1(\lambda_0) k_y]^2}} 
\left[
\begin{array}{c} 
k_x \\
a_1(\lambda_0)k_y
\end{array} \right]\ , \\
\label{V}
&& \mbox{\boldmath$V$}_G^{\gamma}(\lambda_0, \mbox{\boldmath$k$}) =  \frac{1}{\hbar} \, \left[ \begin{array}{c} 
\pr/\pr k_x \\
\pr/\pr k_y
\end{array} \right] 
\varepsilon^{\gamma}_1(\lambda_0, \mbox{\boldmath$k$})
= \frac{\gamma\,v_F}{\sqrt{k_x^2 + [a_1(\lambda_0) k_y]^2}}
\left[
\begin{array}{c} 
k_x \\
a^2_1(\lambda_0)k_y
\end{array} 
\right] \, .
\end{eqnarray}
Here, the spinor vector and its angle with the $x$-axis in Eq.\,\eqref{ts0} represent  crucial components of the  wave functions presented in Eqs.\,\eqref{wfdice1F} and \eqref{wfdice0}, while $\mbox{\boldmath$V$}_G^{\gamma}$ specifies the actual incidence direction of incoming particles and is used to distinguish transmitted and reflected waves associated with the second interface of a barrier layer.
From Eqs.\,\eqref{SV} and \eqref{V}, the angles of these two vectors relative to the $x$-axis are determined by       

\begin{eqnarray}
\label{SVa}
&& \tan \Theta_{\bf S}(\lambda_0) = \left(\frac{k_y}{k_x}\right)\,a_1(\lambda_0) = a_1(\lambda_0) \, \tan \theta_{\bf k}\ , \\
\nonumber
&& \tan \Theta_{\bf V}(\lambda_0) = \left(\frac{k_y}{k_x}\right)\,a_1^2(\lambda_0) = a_1^2(\lambda_0) \, \tan \theta_{\bf k}\ ,
\end{eqnarray}
which will be utilized in our hereafter computations. 
\medskip

As seen in Fig.\,\ref{FIG:3}, direction of the polarization for an imposed dressing field usually mismatches the normal direction of a barrier layer in tunneling structures. As a result, it is reasonable to introduce two coordinate frames of reference  corresponding to these two directions, e.g., $\{x,y\}$ for the $\hat{\mbox{\boldmath$x$}}$ vector whereas $\{x',y'\}$ for the $\hat{\mbox{\boldmath$x$}}'$ vector. These two frames can be related to each other 
by an in-plane rotation angle $\beta$ with a rotation matrix $\hat{\mbb{R}}(\beta)$ given by

\begin{equation}
\hat{\mbb{R}}(\beta) = \left[
\begin{array}{cc}
\cos \beta & -\sin \beta \\
\sin \beta & \cos \beta
\end{array}
\right] \ , 
\label{rota}
\end{equation}
where $\beta$ is the polarization angle of light made with the $\hat{\mbox{\boldmath$x$}}$ direction. From a physics perspective, what makes our transmission problem unique and a lot more complicated results from the fact that both $\{x,y\}$ and $\{x',y'\}$ frames are required to define the  wave function of electrons. First, components of incident-electron momenta outside the barrier layer should be defined in the $(x',y')$ frame with respect to the normal direction of a barrier layer, and meanswhile,  both spinor and group-velocity angles must be calculated in the $(x,y)$ frame within which the energy dispersions can be determined by Eq.\,\eqref{linD}.  Similarly, for electrons within the barrier layer, we also deal with both frames simultaneously as we decide the electron momentum components based on $\mc{E}_0 - V_0 = \hbar v_F \sqrt{\left[k_x^{(2)}\right]^2 + \left[a_1(\lambda_0)\,k_y^{(2)} \right]^2}$ in the $(x,y)$-frame, and the conservation of transverse $k_y'$ component in the $(x',y')$-frame, as illustrated by Fig.\,\ref{FIG:3}.   
\medskip

Another intriguing feature for electron tunneling in a dice lattice with anisotropic dispersions comes from the requirement of a new type of boundary
conditions which are different from simply matching the components of a wave function at two boundaries of a barrier region for graphene and isotropic dice lattice.\,\cite{alphaDice}  In our current system, the new boundary conditions possess an additional $k_{x'}$-related term due to an extra discontinuity at two boundaries of a barrier layer in the $(x',y')$-frame and that the main Hamiltonian in Eq.\,\eqref{HamG} is defined in the $(x,y)$-frame. As a whole, after a lengthy calculation, we find the resulting boundary conditions for a dice lattice are given by $\varphi_2(-\delta x') = \varphi_2(\delta x')$ and 

\begin{equation}
\aleph_\tau^+(\lambda_0, \beta) \, \varphi_1(- \delta x') +  \aleph_\tau^-(\lambda_0, \beta) \, \varphi_3(- \delta x') =
\aleph_\tau^+(\lambda_0, \beta) \, \varphi_1(\delta x') + \aleph_\tau^-(\lambda_0, \beta) \, \varphi_3(\delta x') \ ,
\end{equation}
where $\varphi_j(x)$ for $j=1,\,2,\,3$ represent three wave function components, and 
 
\begin{equation}
\aleph_\tau^\pm (\lambda_0, \beta) = \tau\cos\beta\left[1\pm i\tau a_1(\lambda_0)\tan \beta \right] \ .
\end{equation} 
It is known that a skew or non-symmetric type of Klein paradox is associated with a full transmission independent of barrier height $V_0$, width $W_B$ or 
the energy ${\cal E}_0$ of incoming particles. Such a phenomenon can be observed at a finite incidence angle $\Theta_{V}' \neq 0$, in contrast with the conventional Klein paradox with $\Theta_{V}'=0$. 
In this paper, however, we are interested in investigating 
how this finite angle $\Theta_{V}' \neq 0$ depends on parameters of our system setup, e.g., anisotropy factor $a_1(\lambda_0)$ and light-polarization angle $\beta$ made with $x'$-axis or the potential barrier normal direction. 
\medskip

Common Klein paradox could be observed only if the transverse component of electron momentum vanishes, or equivalently $k_y' = 0$. Obviously, this condition could be met as  
$\theta_{\bf k}' = 0$ or $\theta_{\bf k} = \beta$ and $\mbox{\boldmath$k$}= \{k \cos \beta, k \sin \beta\}$. In fact, this critical incidence angle could 
be found from $\partial\mc{T}[E_{\rm dc},\mc{E}^{\,\gamma = 1}_1(k),\Theta_{{\bf V}'}^{(1)}\,\vert \,\lambda_0,\beta]/\partial\Theta_{{\bf V}'}^{(1)}=0$, which gives rise to\,\cite{LiuM}

\begin{equation}
\label{angleKlein}
\Theta_{{\bf V}'}^{(1)} =\Theta_{{\bf V}}^{(1)} - \beta = \tan^{-1} \left\{
\frac{\left[ a_1^2(\lambda_0) - 1 \right] \tan \beta}{ 1 + a^2_1(\lambda_0) \tan^2\beta}
\, \right\} \ . 
\end{equation}
Consequently, we immediately discern that for $\vert\beta\vert < \pi/2$, the largest asymmetry of the Klein paradox is achieved if $\beta_c = \cot^{-1} [ a_1(\lambda_0) ]$, yielding

\begin{equation}
\left[\Theta_{{\bf V}'}^{(1)}\right]_{\rm max} = \tan^{-1}[a_1(\lambda_0)] - \cot^{-1}[a_1(\lambda_0) ] \ .  
\end{equation}
Therefore, the reach to $\left[\Theta_{{\bf V}'}^{(1)}\right]_{\rm max}$ could be fulfilled by adjusting either anisotropy factor $a_1(\lambda_0)$ or misalignment angle $\beta$. 
\medskip

\begin{figure} 
\centering
\includegraphics[width=0.65\textwidth]{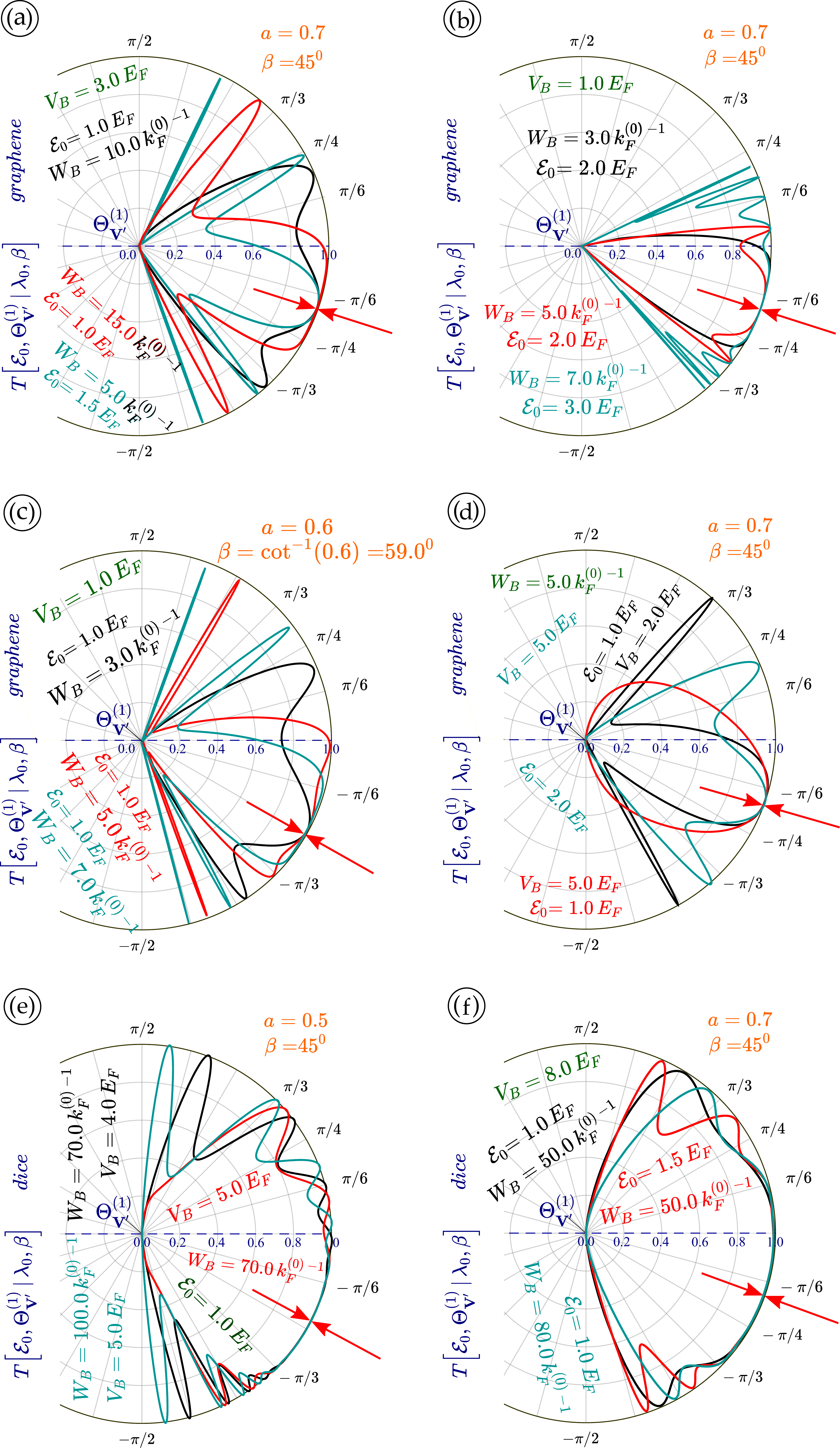}
\caption{(Color online) Angular plots for electron transmission $\mc{T}[\mc{E}_0,\Theta_{{\bf V}'}^{(1)}\,\vert \,\lambda_0,\beta]$ as a function of incidence angle $\Theta_{{\bf V}'}^{(1)}$ of incoming-particle group-velocity vector in graphene and a dice lattice. 
Each panel corresponds to a different value of the anisotropy factor $a_\alpha(\lambda_0)$ of irradiated 
dressed states as well as the misalignment angle $\beta$ and other parameters, as labeled. The Klein paradox is detected as  full transmission is observed for 
different selected values of barrier height $V_0$, incidence energy $\mathcal{E}_0$ and barrier width $W_B$, e.g. $V_0/E_F=3.0$, $\mathcal{E}_0/E_F = 1.0,\,1.5$ and $k_F^{(0)}W_B= 10.0,\,15.0,\,5.0$ for black, red and blue curves, respectively, in panel $(a)$.}
\label{FIG:4}
\end{figure}

\begin{figure} 
\centering
\includegraphics[width=0.65\textwidth]{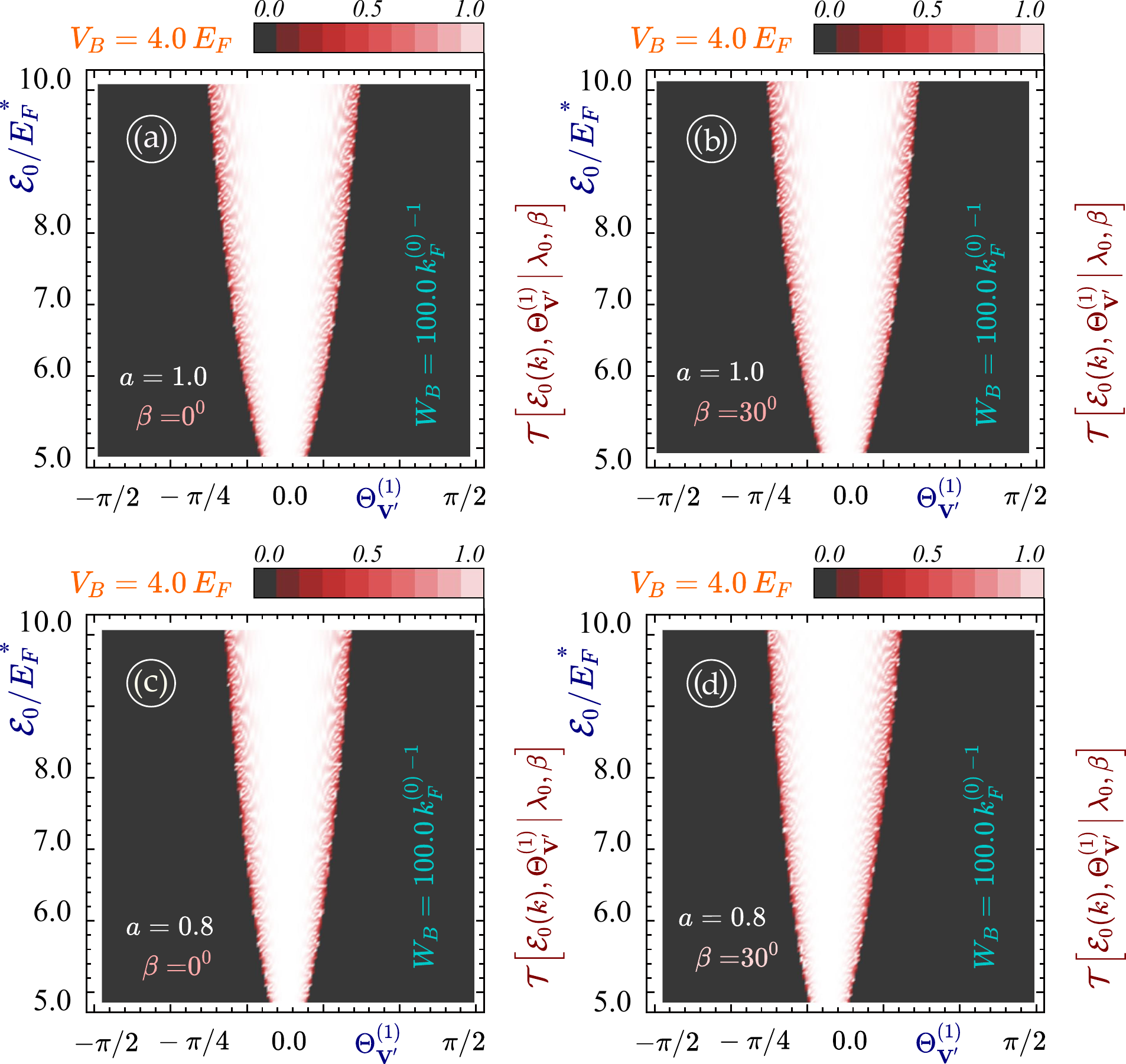}
\caption{(Color online) Density plots for $\mc{T}[\mc{E}_0,\Theta_{{\bf V}'}^{(1)}\,\vert \,\lambda_0,\beta]$ as functions of $\Theta_{{\bf V}'}^{(1)}$ and $\mc{E}_0$ in a dice lattice. Each panel corresponds to a specific value of $a_1(\lambda_0)$ and $\beta$, as labeled. Here $k_F^{(0)}W_B = 100.0$ for all panels. The values of $\mc{E}_0/E_F$ are chosen above the barrier height $V_0/E_F= 4.0$ such that $e$-$h$ (electron-to-hole) transition never occurs in the barrier layer.}
\label{FIG:5}
\end{figure}

\begin{figure} 
\centering
\includegraphics[width=0.65\textwidth]{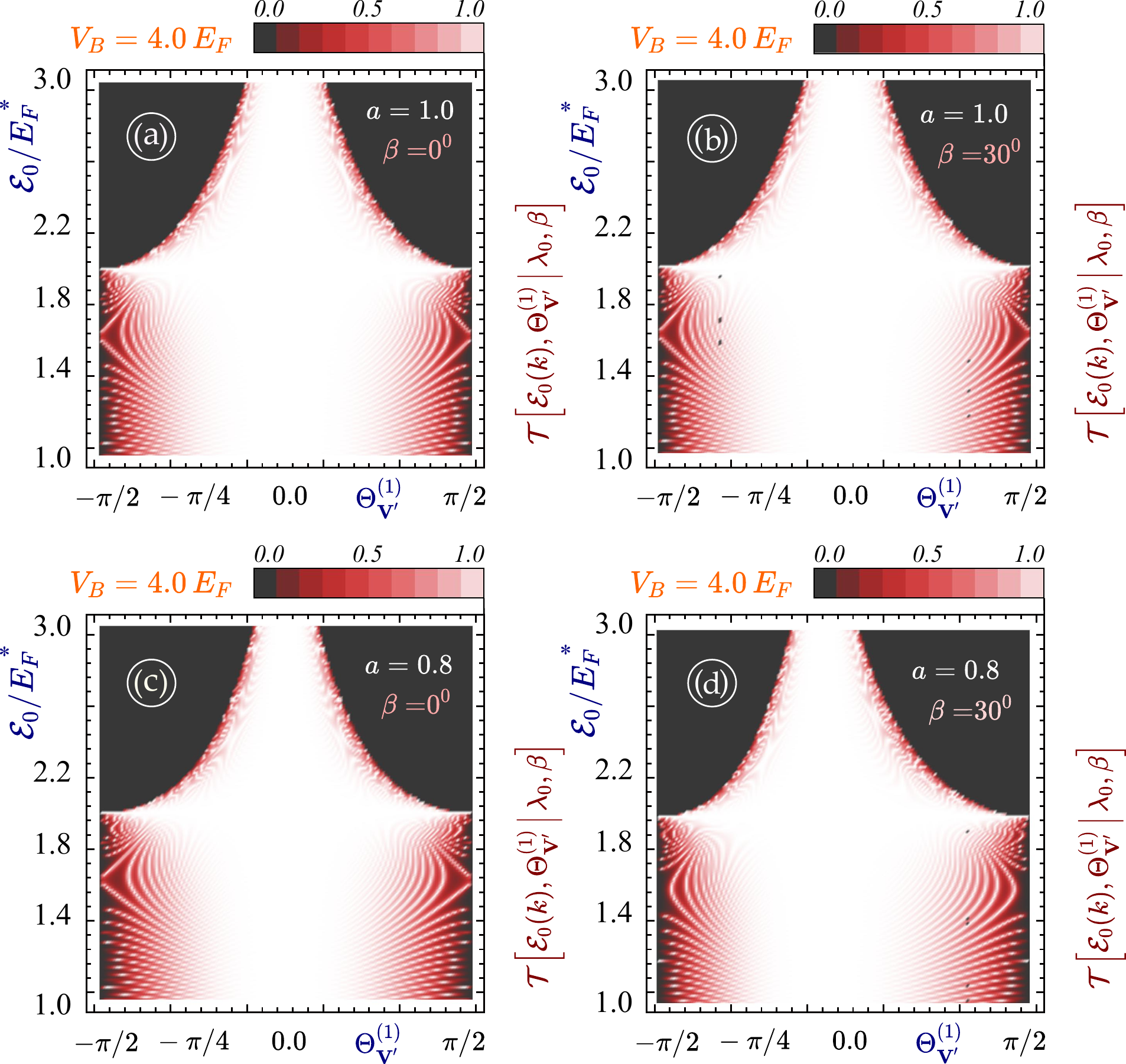}
\caption{(Color online) Density plots for $\mc{T}[\mc{E}_0,\Theta_{{\bf V}'}^{(1)}\,\vert \,\lambda_0,\beta]$ as functions of $\Theta_{{\bf V}'}^{(1)}$ and $\mc{E}_0$ in a dice lattice. Each panel corresponds to a specific value of $a_1(\lambda_0)$ and $\beta$, as labeled. Here, The $k_F^{(0)}W_B = 100.0$ for all panels. The values of $\mc{E}_0/E_F$ are chosen below the barrier height $V_0/E_F = 4.0$ such that $e$-$h$ transition always occurs in the barrier region.}
\label{FIG:6}
\end{figure}

\begin{figure} 
\centering
\includegraphics[width=0.65\textwidth]{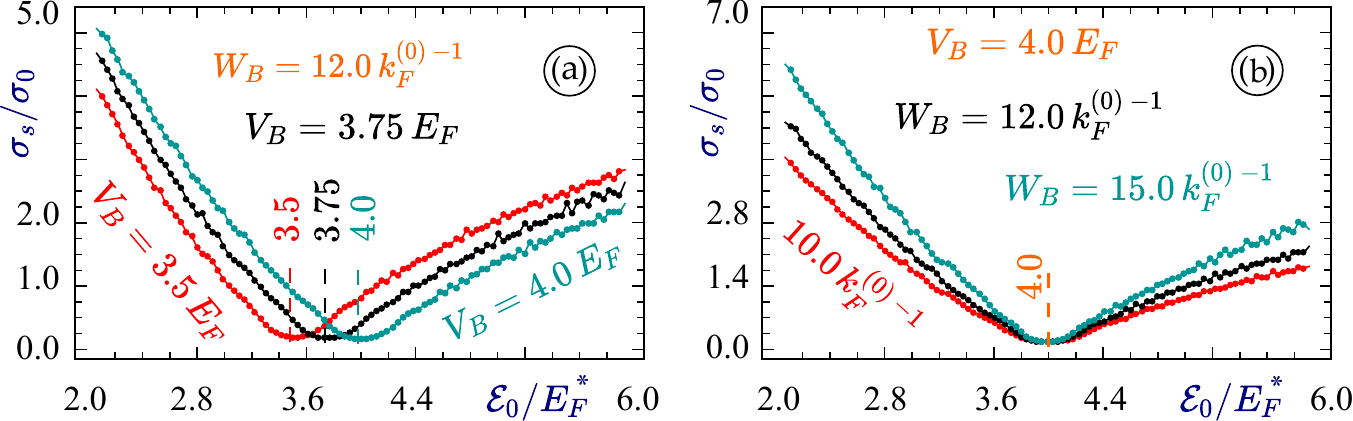}
\caption{(Color online) Tunneling conductivity $\sigma_s(E^*_F,\lambda_0)$ for $\lambda_0=0$ in the units of $\sigma_0 = 2 e^2/h$ for isotropic graphene electronic 
states in the absence of a dressing field as a function of incoming-particle energy $\mc{E}_0=E^*_F$ in the units of $E_F$. Panel $(a)$ is plotted for three different values of barrier height $V_0$ and the fixed barrier width $k_F^{(0)}W_B = 12.0$, while panel $(b)$ displays $\sigma_s(E^*_F,\lambda_0)$ for three different values of $W_B$, as labeled, for fixed $V_0$ and $W_B$.}
\label{FIG:7}
\end{figure}

\begin{figure} 
\centering
\includegraphics[width=0.65\textwidth]{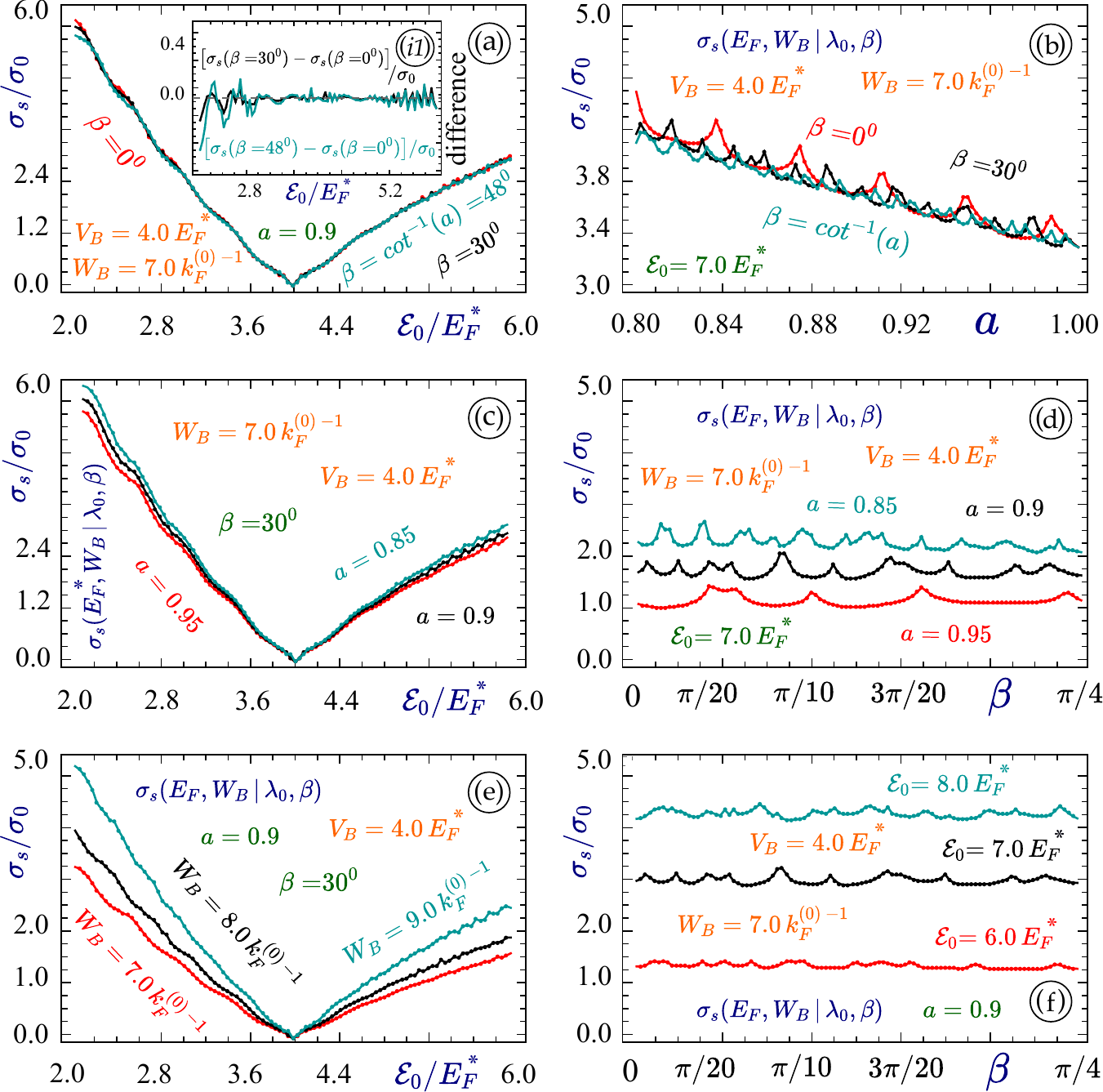}
\caption{(Color online) $\sigma_s(E^*_F,\lambda_0,\beta)$ with $\lambda_0=0$ in the units of $\sigma_0 = 2 e^2/h$ for graphene anisotropic
dressed electronic states. On the left panels $(a)$, $(c)$ and $(e)$, $\sigma_s(E^*_F,\lambda_0,\beta)$ is plotted as a function of $\mc{E}_0=E^*_F$ in the units of $E_F$, as a function of anisotropic factor $a_0(\lambda_0)$ 
in panel $(b)$ and a function of misalignment angle $\beta$ in panels $(d)$ and $(f)$. Inset $(i1)$ in panel $(a)$ displays the difference of tunneling conductivities
with $\beta=30^{\rm o}$ and $\beta = 0$.}
\label{FIG:8}
\end{figure}

\begin{figure} 
\centering
\includegraphics[width=0.65\textwidth]{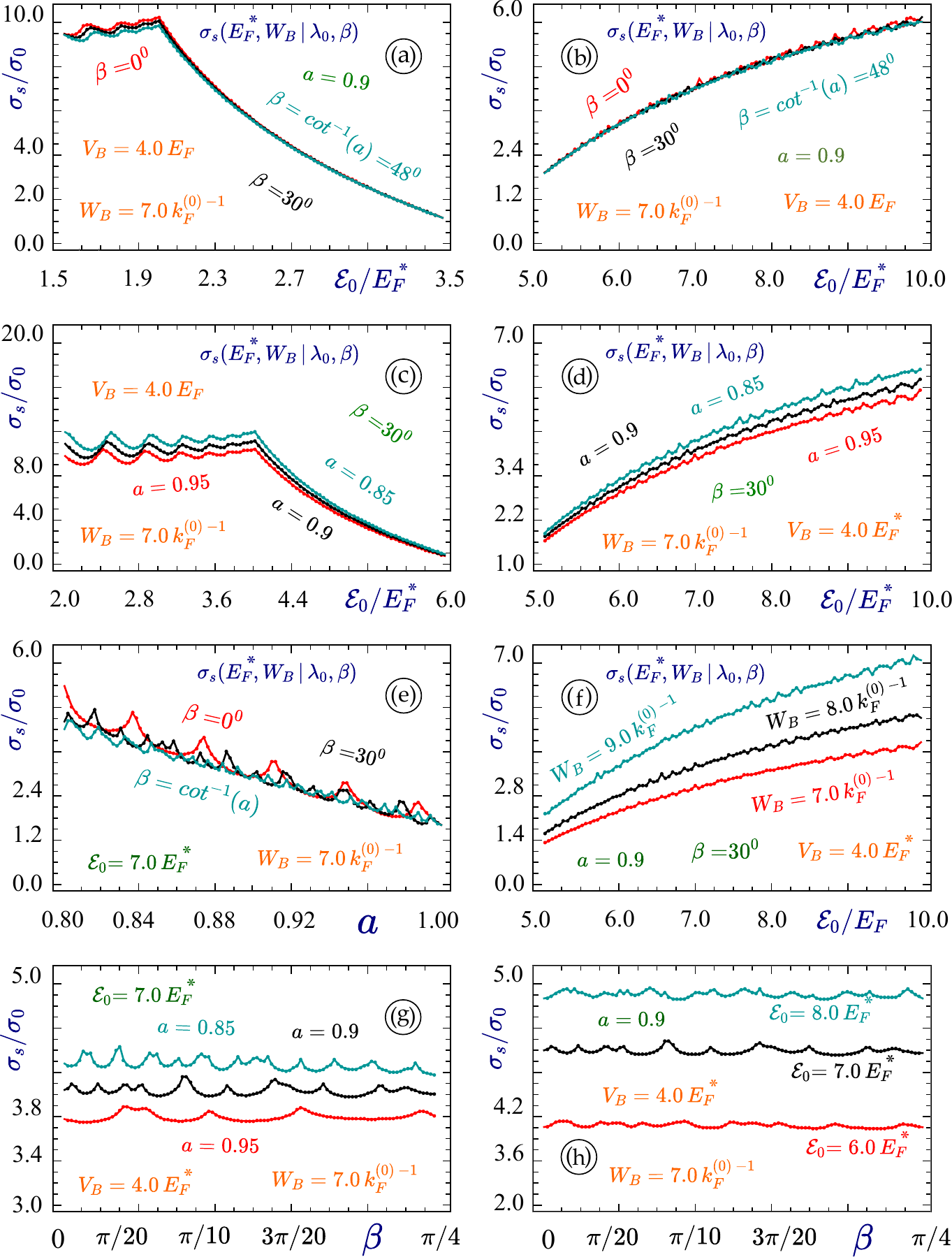}
\caption{(Color online) $\sigma_s(E^*_F,\lambda_0,\beta)$ in the units of $\sigma_0 = 2 e^2/h$ for dice-lattice anisotropic dressed
electron states. On the panels $(a)$-$(d)$ and $(f)$, $\sigma_s(E^*_F,\lambda_0,\beta)$ is plotted as a function of $\mc{E}_0=E_F^*$ in the units of $E_F$, 
as a function of $a_1(\lambda_0)$ in panel $(e)$ and as a function of $\beta$ 
in panels $(g)$ and $(h)$. Here, for $\mc{E}_0=E_F^*$ dependence in isotropic $\sigma_s(E^*_F \vert \lambda_0=0)$, we separate the cases for $\mc{E}_0 > V_0$ and $\mc{E}_0 < V_0$ in the barrier region. }
\label{FIG:9}
\end{figure}	

Our calculated results for electron transmission in both graphene and dice lattice are presented in Figs.\,\ref{FIG:4}, \ref{FIG:5} and \ref{FIG:6}, which clearly demonstrate asymmetric Klein paradox for a finite  $\Theta_{{\bf V}'}^{(1)}$ angle of electron incidence as the anisotropy factor $a_1(\lambda_0)\neq 1$ and the misalignment angle $\beta\neq 0$ between the $(x,y)$ and $(x',y')$ frames.   For the case of a dice lattice, however, we need consider the energy $\mc{E}_0$ of incoming particles separately in the ranges of $\mc{E}_0 < V_0$ and $\mc{E}_0 > V_0$   so as to exclude the possibility for particle scattering into a zero-energy state in the barrier region with an infinite degeneracy for its momentum.  The observed transmission is generally larger for a dice lattice and very close to unity for a wide range of incident angles, which is not the case for graphene with a series of separated distinguished peaks. The dice lattice is also known for its ``magic case'', i.e., a complete transmission for all angles of incidence if $\mc{E}_0=2V_0$ is satisfied. The angular dependence of transmission is not symmetric with respect to $\Theta_{{\bf V}'}^{(1)}=0$ if both $\beta$ and $a_1(\lambda_0)$ are finite [see panel $(d)$ of Figs.\,\ref{FIG:5} and \ref{FIG:6}]. In these cases, the Klein paradox line for  full transmission is moved to negative values for $\Theta_{{\bf V}'}^{(1)}$ in accordance with Eq.\,\eqref{angleKlein}.

\section{Sequential-tunneling current and conductivity}
\label{sec3}

Under a DC electric field $E_{\rm dc}$ along the $x'$ direction for tunneling transport of electrons within monolayer graphene (with $\alpha=0$), which includes a potential barrier distributed along the $x'$ direction with a barrier width $W_B$ and a barrier height $V_0$, the sequential-tunneling sheet current $J_s(E_{\rm dc} \, \vert \,E_F,\lambda_0,\beta)$ in this system is calculated as  

\begin{eqnarray}
\nonumber
J_s(E_{\rm dc} \, \vert \,E_F,\lambda_0,\beta) &=& \frac{g_s g_v\,e}{(2\pi)^2} \, \int d^2\mbox{\boldmath$k$}\,
\mc{T}[\mc{E}^{\,\gamma = 1}_0(\mbox{\boldmath$k$}),\Theta_{{\bf V}'}^{(1)}\,\vert \,\lambda_0,\beta] \, 
V_{G,x'}^{\,\gamma = 1} (\lambda_0, \mbox{\boldmath$k$})\, \\
\label{trans1}
&\times&\left\{ 
f_0[\mc{E}^{\,\gamma = 1}_0(\mbox{\boldmath$k$})] - f_0[\mc{E}^{\,\gamma = 1}_0(\mbox{\boldmath$k$}) + eE_{\rm dc} W_B]
\right\}\ , 
\end{eqnarray}
where $\mc{T}[\mc{E}^{\,\gamma = 1}_0(\mbox{\boldmath$k$}),\Theta_{{\bf V}'}^{(1)}\,\vert \,\lambda_0,\beta]$, which is independent of $E_{\rm dc}$ for a thin barrier layer, 
represents the coefficient for sequential tunneling of electrons through the square-potential barrier,
$g_s=2$ and $g_v=2$ are spin and valley degeneracies of graphene sheet, and 
$\mbox{\boldmath$k$}$ is the wave vector of electrons within the 2D lattice. 
Additionally, $f_0(x)=\{1+\exp[(x-\mu)/k_BT]\}^{-1}$ is the Fermi function for thermal-equilibrium electrons, 
$\mu$ and $T$ are the graphene chemical potential and the system temperature, and the incident energy 
$\mc{E}^{\,\gamma = 1}_0(\mbox{\boldmath$k$})$ is below the Fermi energy of electrons in the source electrode. 
\medskip
\par
For a thin barrier with $eE_{\rm dc} W_B\ll\mc{E}^{\,\gamma = 1}_0(\mbox{\boldmath$k$})$, we find from Eq.\,\eqref{trans1} that

\begin{equation}
 f_0[\mc{E}^{\,\gamma = 1}_0(\mbox{\boldmath$k$})] - f_0[\mc{E}^{\,\gamma = 1}_0(\mbox{\boldmath$k$}) + e E_{\rm dc} W_B] = 
e E_{\rm dc} W_B \, \left[
-\frac{\pr}{\pr \mc{E}^{\,\gamma=1}_0(\mbox{\boldmath$k$})} \,   f_0[\mc{E}^{\,\gamma = 1}_0(\mbox{\boldmath$k$})]
\right] \backsimeq e E_{\rm dc} W_B \,\delta[\mc{E}^{\,\gamma = 1}_0(\mbox{\boldmath$k$})-E^*_F] \ ,
\label{trans2}
\end{equation}
where $\mc{E}^{\,\gamma = 1}_0(\mbox{\boldmath$k$})=\hbar v_F\sqrt{k^2_x+a_0^2(\lambda_0)k_y^2}$,
$E^*_F$ stands for the Fermi energy of graphene layer at low temperatures $k_BT\ll E^*_F$, and $v_F$ is the Fermi velocity. 
In our computations of Eq.\,\eqref{trans1}, we have used the following relations:

\begin{eqnarray}
\label{dtheta}
&& \int d^2\mbox{\boldmath$k$}  = \int\limits_0^{\infty} k \, d k 
\int\limits_0^{2 \pi} d \theta_{\bf k} \ , \\
\label{dtheta1}
&& \theta_{\bf k} = \tan^{-1} \left\{
\frac{1}{a^2_0(\lambda_0)} \, \tan \left(
\Theta_{{\bf V}'}^{(1)} + \beta
\right) \,
\right\}\ ,  \\
\label{dtheta2}
&& d \theta_{\bf k} = a^2_0(\lambda_0) \, \left\{
\cos^2 \left(
\Theta_{{\bf V}'}^{(1)}+\beta
\right) \left[
a^2_0(\lambda_0) + \tan^2
 \left(
\Theta_{{\bf V}'}^{(1)}+\beta
\right)
\right]
\right\}^{-1}  \, d \Theta_{{\bf V}'}^{(1)}\ . 
\end{eqnarray}
Meanwhile, we also acquire

\begin{equation}
\label{Vp}
\left[
\begin{array}{c}
V_{G, x'}^{\gamma} (\lambda_0, \mbox{\boldmath$k$}) \\
V_{G, y'}^{\gamma} (\lambda_0, \mbox{\boldmath$k$})
\end{array}  
\right] =
\frac{\gamma\,v_F}{\sqrt{k_{x'}^2 + a^2_0(\lambda_0)\,k_{y'}^2}}\,
\left[
\begin{array}{c} 
k_{x'} \\
a^2_0(\lambda_0) \, k_{y'}
\end{array} 
\right]=\frac{\gamma\,v_F}{\sqrt{k_{x}^2 + a^2_0(\lambda_0)\,k_{y}^2}}\,\hat{\mbb{R}}(-\beta)\left[
\begin{array}{c} 
k_{x} \\
a^2_0(\lambda_0) \, k_{y}
\end{array} 
\right]  \ ,
\end{equation}
or simply $V_{G, x'}^{\gamma=1} (\lambda_0, \mbox{\boldmath$k$})\equiv v_F\cos\left(
\Theta_{{\bf V}'}^{(1)}\right)$. 
\medskip

Finally, the optically-modulated sheet conductivity $\sigma_s(E^*_F,\lambda_0,\beta)$ of the system is given by

\begin{eqnarray}
\nonumber
\sigma_s(E^*_F,\lambda_0,\beta)&\equiv&\frac{J_s(E_{\rm dc}\, \vert \,E^*_F,\lambda_0,\beta)}{E_{\rm dc}}
=\frac{e^2W_B}{\pi^2}\int\limits_0^\infty k\,dk\int\limits_0^{2\pi} d\theta_{{\bf k}}\,
\mc{T}[\mc{E}^{\,\gamma = 1}_0(\mbox{\boldmath$k$}),\Theta_{{\bf V}'}^{(1)}\,\vert \,\lambda_0,\beta] \, \\
\label{trans3}
&\times&V_{G,x'}^{\,\gamma = 1} (\lambda_0, \mbox{\boldmath$k$})\,\delta[\mc{E}^{\,\gamma = 1}_0(\mbox{\boldmath$k$})-E^*_F]\ . 
\end{eqnarray}
In Eq.\,\eqref{trans3}, the delta function $\delta[\mc{E}^{\,\gamma = 1}_0(\mbox{\boldmath$k$})-E_F]$ for the anisotropic dispersions $\mc{E}^{\,\gamma = 1}_0(\mbox{\boldmath$k$}) = 
\hbar v_Fk f_0(\theta_{\bf k})$ with $f_0(\theta_{\bf k}) = \sqrt{1 -  [1 - a_0^2(\lambda_0)] \, \sin^2(\theta_{\bf k})}$
and $a_0(\lambda_0) < 1$ could be written as 

\begin{equation}
\label{trans4}
\delta[\mc{E}^{\,\gamma = 1}_0(\mbox{\boldmath$k$})-E^*_F] = \left\{ 
 \frac{\pr}{\pr k} \,\left[
\mc{E}^{\,\gamma = 1}_0(\mbox{\boldmath$k$})-E^*_F
\right]
\right\}^{-1} \, \delta[k - k^*_F(\theta_{\bf k})]  
= \frac{1}{\hbar v_F f_0(\theta_{\bf k})} \, \delta[k - k^*_F(\theta_{\bf k})] \ ,
\end{equation}
where $k^*_F(\theta_{\bf k}) = E^*_F/[\hbar v_F f_0(\theta_{\bf k})]$ is the $\theta_{\bf k}$-dependent Fermi momentum for an elliptical 
Dirac-cone dispersion. By using the result in Eq.\,\eqref{trans4} and considering a forward incidence of particles with $|\Theta_{{\bf V}'}^{(1)}|\leq \pi/2$, the conductivity in Eq.\,\eqref{trans3} can be 
calculated explicitly as 

\begin{eqnarray}
\nonumber
&&\sigma_s(E^*_F,\lambda_0,\beta) = \frac{e^2W_BE^*_F}{\pi^2\hbar^2v_F}\int\limits_{-\pi/2}^{\pi/2} \, 
d \Theta_{{\bf V}'}^{(1)}  \, \cos\left(
\Theta_{{\bf V}'}^{(1)}\right)\,
\mc{T}[E^*_F,\Theta_{{\bf V}'}^{(1)}\,\vert \,\lambda_0,\beta]\\
\nonumber
&\times&
\left[\frac{
1 + \tan^2
 \left(
\Theta_{{\bf V}'}^{(1)}+\beta
\right)
}{
a^2_0(\lambda_0) + \tan^2
 \left(
\Theta_{{\bf V}'}^{(1)}+\beta
\right)/a_0^2(\lambda_0)
}\right]
\left[\frac{
	1 + \tan^2
	\left(
	\Theta_{{\bf V}'}^{(1)}+\beta
	\right)/a_0^4(\lambda_0)
}{
	1 + \tan^2
	\left(
	\Theta_{{\bf V}'}^{(1)}+\beta
	\right)/a_0^2(\lambda_0)
}\right]\\
&=&\left(\frac{2e^2}{h}\right)\frac{E^*_FW_B}{\hbar v_F}\int\limits_{-\pi/2}^{\pi/2} \, 
\frac{d\Theta_{{\bf V}'}^{(1)}}{\pi}  \, \cos\left(
\Theta_{{\bf V}'}^{(1)}\right)\,
\mc{T}[E^*_F,\Theta_{{\bf V}'}^{(1)}\,\vert \,\lambda_0,\beta]
\left[\frac{
	1 + \tan^2
	\left(
	\Theta_{{\bf V}'}^{(1)}+\beta
	\right)
}{
	a_0^2(\lambda_0) + \tan^2
	\left(
	\Theta_{{\bf V}'}^{(1)}+\beta
	\right)
}\right]\ ,
\label{sigma0}
\end{eqnarray}
where $\mc{E}^{\gamma=1}_0(\mbox{\boldmath$k$})=E^*_F$ represents the kinetic energy of tunneling electrons at the Fermi level of graphene. 
\medskip 
 
Finalizing our calculations, we also obtain the density-of-states for asymmetric Dirac-cone electrons in graphene, as well as the dependence of their Fermi energy $E^*_F$ on the electron density $n_e$\,\cite{huang2018many}. In general, the density-of-states $\rho_S (\mbb{E})$ is defined as 

\begin{equation}
\label{dos}
\rho_S (\mbb{E}) = \int \, \frac{d^2\mbox{\boldmath$k$}}{(2 \pi)^2} \, \sum\limits_{\gamma=\pm 1} \, \sum\limits_{\xi, \sigma = \pm 1}  \, 
\delta\left[ \mbb{E} - \mathcal{E}_0^{\gamma} (k) \right]  \, , 
\end{equation} 
where $\sigma$ and $\xi$ refer to the spin and valley indices.  
As an example, we will only consider electron states with $\gamma = 1$, while the situation for hole states with $\gamma=-1$ will be exactly symmetric. 
Meanwhile, there exist valley and spin degeneracies so that $ \sum\limits_{\xi,\sigma = \pm 1}=4$. 
Moreover, Eq.\,\eqref{dos} looks identical to a part of the integrand in Eq.\,\eqref{sigma0} for conductivity, therefore, we can easily rewrite Eq.\,\eqref{dos} as

\begin{equation}
\label{dos2}
\rho_S (\mbb{E}) = \mathcal{C}_{\lambda_0}\, \mathbb{E} \equiv\frac{\mbb{E}}{(\pi \hbar v_F)^2} \, \int\limits_{0}^{2 \pi} \, d \theta_{\bf k} \,
\left[\frac{
  1 +   \tan^2 \theta_{{\bf k}}
}{
	1 +  a^2_0(\lambda_0) \tan^2\theta_{{\bf k}}}\right] \ ,
\end{equation}  
which is still proportional to energy $\mbb{E}$ although its coefficient depends on the anisotropy factor $a_0(\lambda_0) < 1$, and  
is reduced with increasing light intensity, in correspondence with the increase of tunneling current due to the anisotropic dispersion. 
Furthermore, the Fermi energy $E_F$ of the system is calculated as $E^*_F = (2 n_e / \mathcal{C}_{\lambda_0})^{1/2}$ which reduces to a standard expression $E^*_F = \hbar v_F \,\sqrt{\pi n_e}$ after taking $a_0(\lambda_0) = 1$. 

\medskip

The tunneling conductivity $\sigma_s(E^*_F,\lambda_0)$ for the standard case of isotropic dispersions $(\lambda_0=0)$ and symmetric Klein paradox $(\Theta_{{\bf V}'}^{(1)}=0)$ is shown in Fig.\,\ref{FIG:7}. As it is well known,  $\sigma_s(E^*_F,\lambda_0)$ reproduces a familiar $V$-shape dependence for $\mc{E}_0=E^*_F$, where the transmission $\mc{T}[\mc{E}_0,\Theta_{{\bf V}'}^{(1)}\,\vert \,\lambda_0]$  is equal to zero for all non-zero $\Theta_{{\bf V}'}^{(1)}$ angles of incidence and reaches its minimal values at $\mc{E}_0=V_0$, where the electron momentum in the barrier region is zero so that the particle still remains. However, the Klein paradox is still observed for the head-on collision.  Similar $\mc{E}_0$ dependence in $\sigma_s(E^*_F,\lambda_0,\beta)$ is found for anisotropic dispersions with a finite angle $\beta$, as seen in Fig.\,\ref{FIG:8}, which is also reflected in $\mc{T}[\mc{E}_0,\Theta_{{\bf V}'}^{(1)}\,\vert \,\lambda_0,\beta]$. It is very interesting to find that the $\beta$ dependence of $\sigma_s(E^*_F,\lambda_0,\beta)$ in Figs.\,\ref{FIG:8}$(d)$ and \ref{FIG:8}$(f)$ is weak, periodic and non-monotonic so that it plays no substantial role in $\sigma_s(E^*_F,\lambda_0,\beta)$. The fluctuations around a fixed value in  Figs.\,\ref{FIG:8}$(d)$ and \ref{FIG:8}$(f)$ result from the fact that sub-resonances with $k_x^{\prime(2)}W_B=n\pi$ with integer $n$\,\cite{neto} can equally contribute to tunneling current and the change of $\beta$ corresponds to a rotation of these ``transmission petals''.  In fact, as presented in Eq.\,\eqref{trans3}, $\sigma_s(E^*_F,\lambda_0,\beta)$ is a convolution of $\mc{T}[\mc{E}_0,\Theta_{{\bf V}'}^{(1)}\,\vert \,\lambda_0,\beta]$  with $\cos \left[ \Theta_{{\bf V}'}^{(1)} \right]$, which certainly relies on the misalignment angle $\beta$.
\medskip

It is also crucial that $\sigma_s(E^*_F,\lambda_0,\beta)$ exhibits a stable decrease with increasing anisotropy factor $a_\alpha(\lambda_0)$, namely, $\sigma_s(E^*_F,\lambda_0,\beta)$ will reach its minimal value  for the isotropic case with $a_\alpha(\lambda_0) = 1$ and demonstrate a steady growth with enhanced eccentricity $e_\alpha(\lambda_0)=\sqrt{1-a^2_\alpha(\lambda_0)}$ of the electron dispersions.  Therefore, the presence of an optical dressing field is enabled to enhance the tunneling conductivity of electrons with increasing field intensity, which demonstrates an attractive device application. This property remains true for both graphene and dice lattices, as can be verified from Fig.\,\ref{FIG:9}$(e)$, and could be explained by a similar dependence of density-of-states in Eq.\,\eqref{dos2}  as well as the dependence of $E^*_F$ on $n_e$ in Eq.\,\eqref{dos2} for $a_\alpha(\lambda_0)<1$.   As illustrated in the right panel of Fig.\,\ref{FIG:3}, the actual value of a transverse wave number $k_y$ will be changed by the inclusion of anisotropy.  

\medskip
\par
Mathematically, the dependence of $\sigma_s(E^*_F,\lambda_0,\beta)$ on the light-induced anisotropy factor $a_\alpha(\lambda_0)$ can be quantified by two competing factors, i.e., a reduced dependence of $\theta_{\bf k}$ on the group velocity angle $\Theta_{{\bf V}'}^{(1)}$ in Eq.\,\eqref{dtheta1} and a slight decrease of density-of-states $\rho_S (\mbb{E})$ included in Eq.\,\eqref{sigma0} due to the interchange of $\mbox{\boldmath$k$}-$integration defined in Eq.\,\eqref{dtheta2}.  On the other hand, the Fermi energy $E_F^*$ is considerably reduced for an anisotropic Dirac cone under a fixed electron density $n_e$, and therefore, it should be regarded as a key parameter in our computations. Moreover, as $a_\alpha(\lambda_0)=1$, the isotropic $\sigma_s(E^*_F \vert \lambda_0=0)$ also displays a steady enhancement for a wider barrier width $W_B$ mainly due to a bigger voltage drop across the barrier layer, as seen in Eq.\,\eqref{sigma0}.   Furthermore, the tunneling current in a dice lattice reveals qualitatively the same feature as in graphene by comparing Fig.\,\ref{FIG:8} with Fig.\,\ref{FIG:9}, but the actual values of $\sigma_s(E^*_F, \vert \lambda_0=0)$ are quite different due to an enhanced transmission coefficient for a dice lattice compared to that of graphene, as can be seen in Fig.\,\ref{FIG:4} and from Ref.\,[\onlinecite{iurov2020klein}]. Meanwhile, we find additional  saturation of current increasing as $\mc{E}_0=E^*_F$ becomes lower than a critical value, as found from Figs.\,\ref{FIG:9}$(a)$ and \ref{FIG:9}$(c)$, due to unique features in the transmission coefficient of a dice lattice.    

\section{Summary and Remarks}
\label{sec4}

The effort of our work has been directed towards  carrying out a rigorous investigation of the tunneling conductance and the currents through a square potential barrier in graphene and a dice lattice in the presence of irradiation-induced anisotropy of the electronic states and, specifically, their energy dispersions. In order to do that, we have derived an analytical expression for the conductivity as a function of all the crucial tuning parameters: an incoming electron energy $\mc{E}_0$, angle of  incidence $\Theta_{{\bf V}'}^{(1)}$, anisotropy $a(\lambda_0)$ and misalignment angle $\beta$ between the directions of the light polarization and the direct incidence on the potential barrier.

\par
For a finite angle $\beta$, the direction of motion of an incoming particle is determined by its group velocity $\vec{V}_{G}^{\gamma} (\lambda_0, {\bf k})$ and its angle $\Theta_{{\bf V}'}^{(1)}$ relative to the normal incidence on the potential barrier, which is also used to distinguish the transmitted and reflected waves in the barrier region, but not the wavevector angle $\theta_{\bf k}$. In our case, however, both these angles enter the equations for conductance simultaneously, which brings in some new unexpected relations and physical behavior.

\par
We have found that the tunneling conductance has only non-essential and periodic dependence on the misalignment angle $\beta$. Its dependence on anisotropy consists of two terms with competing effects: a monotonically increasing trend coming from the mismatch between angles $\Theta_{{\bf V}'}^{(1)}$ and $\theta_{\bf k}$, and a separate decreasing dependence which exactly reflects the way how the density of states for such anisotropic Dirac electrons is related to their Fermi energy.

\par
The external optical dressing field could modify the sequential-tunneling current of electrons with the help of polarization angle $\beta$ and light-interaction strength $\lambda_0$. In addition to the applied $E_{\rm dc}$, the sequential tunneling of electrons can also be enhanced by the Fermi energy $E_F$ in two electrodes and the barrier width $W_B$. A strong dependence of the tunneling conductance and the current on the electron doping and its fine-tuning by the applied irradiation leads to a possibility to engineer an electronic device with desirable characteristics. Therefore, our work is expected to be an important step in understanding the radiation-modulated properties and behavior of such optoelectronic devices.

\section*{Acknowledgement(s)}
G.G. would like to acknowledge the support from the Air Force Research Laboratory (AFRL)
through Grant No. FA9453-21-1-0046. D.H. was supported by the Air Force Office of Scientific Research (AFOSR).


\bibliography{DiceCurrent}

\end{document}